\documentclass[
    prd,
    twocolumn,
    aps,
    superscriptaddress,
    reprint,
]{revtex4-2}

\usepackage[T1]{fontenc}
\usepackage{amsmath,amssymb}
\usepackage{mathtools}
\usepackage[mathcal]{euscript}
\usepackage{microtype}
\usepackage{bm}
\usepackage{subfigure}
\usepackage{makecell}

\usepackage{multirow}
\usepackage{graphicx}
\graphicspath{{figures/}}
\usepackage{wrapfig}

\usepackage[svgnames,table]{xcolor}
\usepackage[
    colorlinks=true,
    citecolor=DimGray,
    linkcolor=SlateBlue,
    urlcolor=Navy,
    ]{hyperref}

\usepackage[capitalise]{cleveref}
\usepackage[
    range-phrase=\text{\textendash},
    range-units=single,
]{siunitx}
\DeclareSIUnit{\year}{yr}
\usepackage{dcolumn}

\usepackage{comment}
\usepackage{natbib}
\usepackage{orcidlink}

\newcommand{\rme}{\mathrm{e}}
\newcommand{\rmi}{\mathrm{i}}

\newcommand{\sechead}[1]{\vspace*{10pt}

\noindent\emph{#1 ---}
}

\begin{document}

\title{Loss-tolerant hybrid metasurface mirror for reducing coating Brownian noise in precision optical cavities}

\author{Swadha Pandey\,\orcidlink{0000-0002-2426-6781}}
\email{swadha@mit.edu}
\author{Evan D. Hall\,\orcidlink{0000-0001-9018-666X}}
\author{Peter Fritschel}
\author{Matthew Evans\,\orcidlink{0000-0001-8459-4499}}
\affiliation{LIGO Laboratory, Department of Physics, Massachusetts Institute of Technology, Cambridge, MA 02139, USA}

\date{\today}

\begin{abstract}
We propose a hybrid mirror design that places an embedded metasurface beneath a Bragg reflector, combining the low coating Brownian noise of metasurface mirrors with the low optical loss of conventional multilayer coatings. Unlike previous hybrid designs, which place the metasurface at the input surface, our design positions it below the Bragg stack, reducing the optical field incident on the loss-prone resonant layer. For a design operating at 1064 nm, we show that this reduces absorption from \qty{24}{ppm} for a bare metasurface to \qty{7}{ppm} for the hybrid. 
We introduce a general method for calculating the coating Brownian noise of hybrid metasurface-multilayer mirrors at arbitrary coupling strength, extending prior methods to coherently combine a beam-scale mean-field contribution with the periodic, sub-wavelength-scale response of the metasurface. Applying this to our design, we find a reduction in Brownian noise amplitude spectral density relative to an equivalent-reflectance Bragg mirror by a factor of 1.6. 
Finally, we show that this design has higher angular tolerance than a bare metasurface mirror, maintaining higher reflectance over an angular range of \qty{1}{\degree}. This work provides a practical path towards incorporating low-noise metasurface mirrors into precision optical cavities, while minimizing optical loss and angular sensitivity.
\end{abstract}

\maketitle

\section{Introduction}
\label{sec:intro}

Coating Brownian noise, caused by mechanical loss in the materials of high-reflectance Bragg mirror coatings, is a limiting noise source in precision physics experiments, from gravitational-wave detectors~\cite{Harry:2002} to atomic clocks~\cite{Numata:2004}. Two main approaches are being pursued to address this limitation. The first is to find and use materials with inherently lower mechanical loss, both amorphous~\cite{Vajente:2021,McGhee:2023} and crystalline~\cite{Penn:2019,Lee:2026}. The second combines materials with complementary properties~--~typically a material with lower mechanical loss and higher optical loss, together with a material with higher mechanical loss and lower optical loss~--~in an optimized layer configuration to capture the benefits of both~\cite{Yam:2015,Steinlechner:2015}. This multi-material approach has drawn increasing attention in recent work~\cite{Tait:2020}.

A third approach uses metasurfaces, which are resonant mirrors formed from a single layer of periodic sub-wavelength nanostructures~\cite{Magnusson:1992,Wang:1993}. High reflectance is achieved through resonant interference via guided-mode resonances supported by the nanostructures, rather than through the constructive interference of many quarter-wavelength layers used in a conventional Bragg stack. This approach is motivated by the fact that metasurfaces can achieve high reflectance using far less coating material, hundreds of nanometers rather than the several micrometers required for high-reflectance Bragg stacks, which in principle could lead to lower overall Brownian noise~\cite{Kroker:2017,Dickmann:2018}. However, single-layer metasurface mirrors are highly sensitive to fabrication errors and prone to high optical loss, and it has proven difficult to experimentally realize their high theoretical reflectance~\cite{Bruckner:2010,Atikian:2022}. Recent work has combined metasurface mirrors with Bragg stacks in what we term a hybrid approach~\cite{Dickmann:2018,Dickmann:2023,Kranhold:2026,Hartmann:2026}, conceptually similar to the multi-material route described above. In these designs, a metasurface mirror is placed on top of a reduced Bragg stack, allowing high total reflectance to be achieved experimentally while retaining the Brownian noise reduction the metasurface provides. However, this does not address the high optical loss that the metasurface structure itself is prone to.

This high optical loss arises from the sensitivity of the resonant structure to fabrication irregularities, which leads to high absorption and scattering. We propose a hybrid design that mitigates the resulting optical loss without altering the fabrication tolerances of the metasurface itself. By embedding the metasurface in glass rather than leaving it air-clad, as shown in \cref{fig:design}, we can position it beneath a reduced Bragg stack. The Bragg stack then decreases the electromagnetic field incident on the metasurface, lowering the amount of optical loss the metasurface introduces while preserving its Brownian-noise benefit, as we show in \cref{sec:noise}.

In \cref{sec:design}, we detail this mirror architecture and present a potential implementation operating at \qty{1064}{\nm}. Using realistic material absorption levels, we show that this design achieves high reflectance with an absorption of \qty{7}{ppm}, compared to \qty{24}{ppm} for a bare metasurface. In \cref{sec:fab}, we briefly outline a possible fabrication process. In \cref{sec:noise}, we present a general methodology for rigorously calculating the Brownian noise of hybrid mirrors with arbitrary coupling strength between the metasurface and Bragg stack, extending existing methods~\cite{Levin:1998} for periodic metasurface structures~\cite{Kroker:2017} and for multilayer coatings~\cite{Hong:2013} to coherently combine their respective contributions, and show that our design reduces the Brownian noise amplitude spectral density by a factor of 1.6 relative to an equivalent-reflectance Bragg mirror. As mirror designs involving metasurfaces gather increasing attention, this general method for calculating their coating Brownian noise will enable a meaningful comparison to other mirror designs, allowing us to realistically assess their advantages. In \cref{sec:sensitivity}, we examine the angular sensitivity of this design, a known limitation of metasurface mirrors, and show that it retains higher reflectance than a bare metasurface mirror over an angular range of \qty{1}{\degree}. We believe that this design is an important step towards incorporating metasurfaces into precision measurement experiments, which impose stringent limits on tolerable optical loss.

\section{Design}
\label{sec:design}

\begin{figure*}[t]
    \centering
    \includegraphics[width=\textwidth]{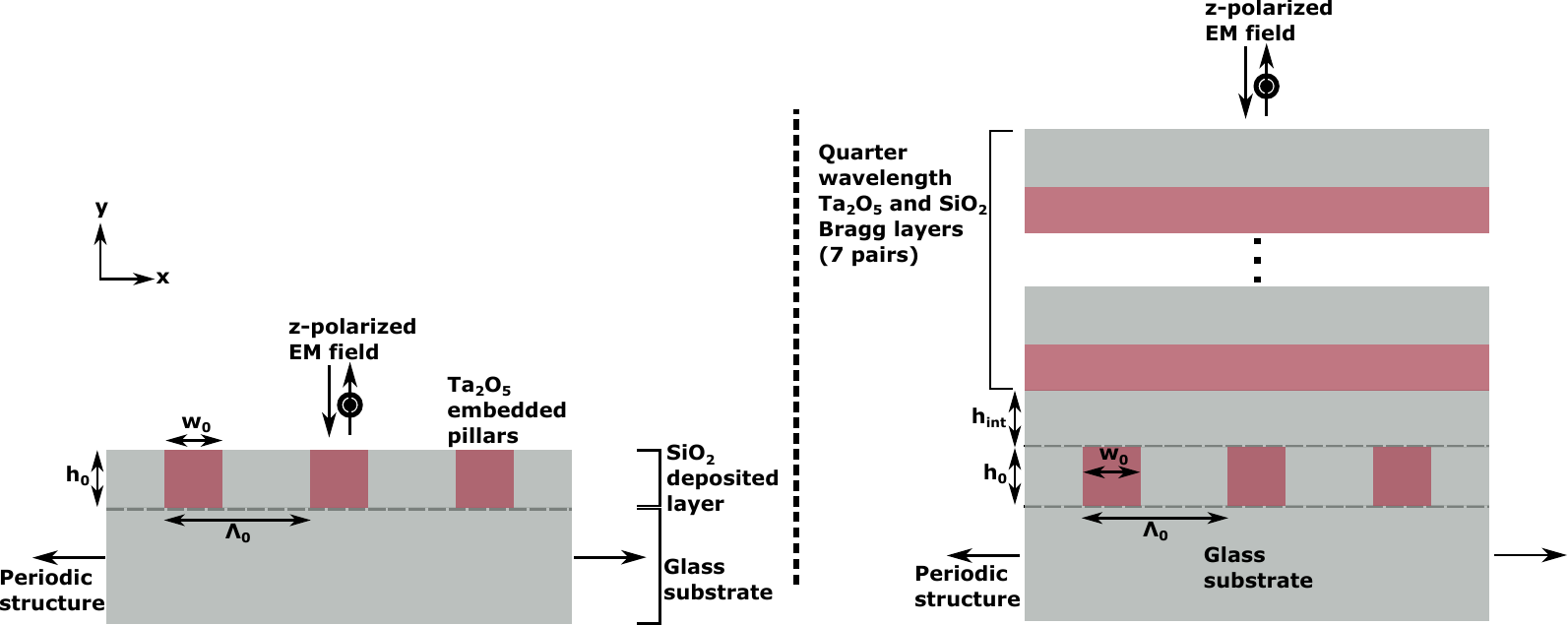}
    \caption{Mirror designs, with parameters optimized to maximize reflectance of transverse electric (TE) polarized light at a wavelength of \qty{1064}{\nm}. \emph{Left:} embedded metasurface mirror, consisting of a one-dimensional array of tantalum pentoxide (Ta$_2$O$_5$) nanopillars of width ($w_0$), height ($h_0$), and periodicity ($\Lambda_0$), on a fused silica substrate. The Ta$_2$O$_5$ nanopillars are embedded in a thin-film silicon dioxide (SiO$_2$) layer of the same height ($h_0$). \emph{Right:} the hybrid mirror design, where the embedded metasurface mirror is buried underneath a Bragg stack consisting of 7 pairs of alternating quarter-wavelength layers of Ta$_2$O$_5$ and SiO$_2$, with an intermediate SiO$_2$ layer of height $h_\mathrm{int}$ for phase matching.}
    \label{fig:design}
\end{figure*}

\begin{table}
    \centering
    \caption{Parameter values for the mirror designed for operation at wavelength \qty{1064}{\nm}: width ($w_0$), height ($h_0$), and periodicity ($\Lambda_0$) of the embedded Ta$_2$O$_5$ pillar, and height of the intermediate SiO$_2$ layer ($h_\mathrm{int}$), as shown in \cref{fig:design}. $n_\mathrm{layers}$ is the total number of alternating quarter-wavelength layers of Ta$_2$O$_5$ and SiO$_2$ above the metasurface.}
    \label{tab:params}
    \begin{tabular}{|c || c | c |} 
        \hline
        Parameter & Value & Unit\\ 
        \hline\hline
        $w_0$ & 282.0 & nm\\
        \hline
        $h_0$ & 305.0 & nm\\ 
        \hline
        $\Lambda_0$ & 704.0 & nm\\
        \hline
        $h_\mathrm{int}$ & 190.2 & nm\\ 
        \hline
        $n_\mathrm{layers}$ & 14 & --\\ 
        \hline
    \end{tabular}
\end{table}

The proposed mirror architecture is shown in \cref{fig:design}. We first consider a single-layer embedded metasurface mirror, consisting of a one-dimensional array of rectangular tantalum pentoxide (Ta$_2$O$_5$) nanopillars of width ($w_0$), height ($h_0$), and periodicity ($\Lambda_0$), on a glass substrate. The pillars are surrounded by a silicon dioxide (SiO$_2$) layer of the same height, $h_0$. As shown in \cref{fig:design}, the structure is periodic in the $\hat{x}$-direction, and the light is incoming in the $-\hat{y}$-direction. We simulate the mirror using rigorous coupled-wave analysis (RCWA)~\cite{Moharam:1981}, and optimize the mirror parameters $w_0$, $h_0$, and $\Lambda_0$, to maximize reflectance for transverse electric (TE) polarized light (along the $\hat{z}$-direction) of wavelength $\lambda_0=$\qty{1064}{\nm} at normal incidence, by exciting a guided-mode resonance~\cite{Magnusson:1992}. The values of the optimized parameters are displayed in \cref{tab:params}. For simplicity, we consider a single polarization and a one-dimensional array, although the design can be made polarization insensitive by using two-dimensional metasurfaces. We have chosen Ta$_2$O$_5$ and SiO$_2$ as our materials, as they are used in current gravitational-wave detectors and have low intrinsic mechanical loss~\cite{Granata:2020}. However, our approach is directly applicable to different materials as well as two-dimensional arrays.

\begin{table}
    \centering
    \caption{Material parameters used in analysis for Ta$_2$O$_5$ and SiO$_2$: extinction coefficient (k), refractive index (n), mechanical loss ($\phi$), Young's modulus (Y), Poisson's ratio ($\nu$). Ion-beam sputtering (IBS) values are taken from Refs.~\cite{gwinc,Flaminio:2010,Granata:2020}. The extinction coefficient (k) is different for different material deposition techniques, IBS for the intermediate and Bragg layers and atomic layer deposition (ALD) for the metasurface.}
    \label{tab:material}
    \begin{tabular}{| c || c | c || c | c || c |}
        \hline
        \multirow{2}{*}{Property} & \multicolumn{2}{c||}{Ta$_2$O$_5$} & \multicolumn{2}{c||}{SiO$_2$} & \multirow{2}{*}{Unit} \\
        \cline{2-5}
         & IBS & ALD & IBS & ALD & \\
        \hline\hline
        $k$   & $5\times10^{-7}$ & $10^{-6}$ & 0 & $10^{-7}$ & -- \\
        \hline
        $n$   & \multicolumn{2}{c||}{2.09} & \multicolumn{2}{c||}{1.45} & -- \\
        \hline
        $\phi$ & \multicolumn{2}{c||}{$3.9\times10^{-4}$} & \multicolumn{2}{c||}{$2.3\times10^{-5}$} & -- \\
        \hline
        $Y$   & \multicolumn{2}{c||}{120} & \multicolumn{2}{c||}{70} & GPa \\
        \hline
        $\nu$ & \multicolumn{2}{c||}{0.29} & \multicolumn{2}{c||}{0.19} & -- \\
        \hline
    \end{tabular}
\end{table}

We then simulate our hybrid mirror architecture, shown in \cref{fig:design}, also using RCWA. The optimized embedded metasurface mirror is placed underneath a reduced Bragg stack consisting of alternating layers of Ta$_2$O$_5$ and SiO$_2$. The number of Bragg layer pairs is a free design parameter~--~increasing it raises the achievable total reflectance and further reduces the field reaching the metasurface, at the cost of additional coating thickness. We use 7 pairs here as an illustrative example. An intermediate SiO$_2$ layer is placed above the metasurface and below the Bragg stack, with height $h_\mathrm{int}$ chosen to phase-match the reflections from the metasurface and Bragg stack so they interfere constructively. $h_\mathrm{int}$ was determined via an RCWA-based parameter sweep to maximize total reflectance. As discussed above, the metasurface layer is prone to optical loss from its sensitivity to fabrication irregularities. Burying it beneath the Bragg stack allows the mirror to achieve high total reflectance while reducing the electromagnetic field that reaches the metasurface, which in turn lowers the overall optical loss of the design. The total height of this hybrid coating is \qty{2.7}{\micro\meter} with 16 layers (14 Bragg layers, one intermediate layer, and one metasurface layer), whereas an equivalent-reflectance Bragg stack composed of the same materials would require 36 layers and a total thickness of \qty{5.8}{\micro\meter}. This reduction in total thickness is expected to reduce coating Brownian noise, as detailed in \cref{sec:noise}.

To illustrate the advantage of this design, we begin by simulating the single-layer bare embedded metasurface mirror with loss-free materials. We then introduce realistic absorption known to occur in thin films deposited via techniques such as atomic layer deposition (ALD), using the refractive index (n) and extinction coefficient (k) values tabulated in \cref{tab:material} for ALD-deposited Ta$_2$O$_5$ and SiO$_2$, applied to both the pillars and the surrounding embedding layer of the metasurface. The extinction coefficient values are extrapolated from our own measurements of ALD-deposited amorphous oxides, on the basis that both materials are oxides deposited under similar process conditions; this estimate is detailed in the Appendix. We emphasize that these extinction coefficient values are intended as order-of-magnitude estimates representative of realistic nanofabrication-induced absorption, in order to phenomenologically study the expected optical loss. We then simulate the full hybrid design, for which we assume ion-beam sputtering (IBS) deposition of the intermediate and quarter-wavelength Bragg layers, the current state-of-the-art for gravitational-wave detector coatings, using the corresponding extinction coefficients~\cite{Granata:2020}, also tabulated in \cref{tab:material}. The substrate is fused silica, for which we use the same optical and mechanical parameters as IBS-deposited SiO$_2$ in \cref{tab:material}.

\begin{figure}[]
    \centering
    \includegraphics[width=\columnwidth]{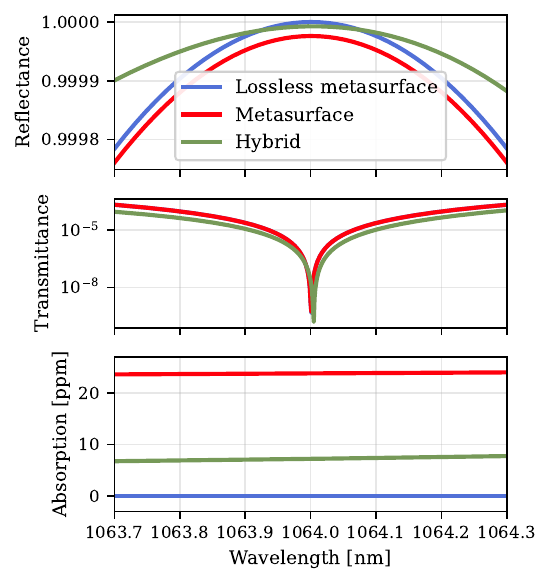}
    \caption{Reflectance (top), transmittance (middle), and absorption (bottom) spectra for the various optimized mirror designs: lossless embedded metasurface (blue), lossy embedded metasurface (red), full hybrid design (green).}
    \label{fig:spectra}
\end{figure}

The resulting reflectance (R), transmittance (T), and absorption (A) spectra are shown in \cref{fig:spectra}. The lossless metasurface achieves high reflectance, which degrades once absorption is introduced into the metasurface materials. The hybrid structure recovers this reflectance and broadens the bandwidth, as the Bragg stack has an inherently broader reflectance band compared to the narrow resonance of the metasurface alone. For the same two designs, the total absorption reduces from \qty{24}{ppm} for the lossy metasurface to \qty{7}{ppm} for the hybrid, matching the reduction in total loss ($1-R$, from \qty{24}{ppm} to \qty{7}{ppm}) and consistent with transmittance remaining negligible in both cases.

We have considered one specific source of optical loss, arising from absorption in the deposited materials. However, the same field-suppression mechanism responsible for reducing absorption is expected to reduce other loss channels as well, such as scatter from surface roughness and sensitivity to nanostructure dimensional errors, since these too scale with the electromagnetic field intensity at the metasurface.

\begin{figure}[]
    \centering
    \includegraphics[width=0.8\columnwidth]{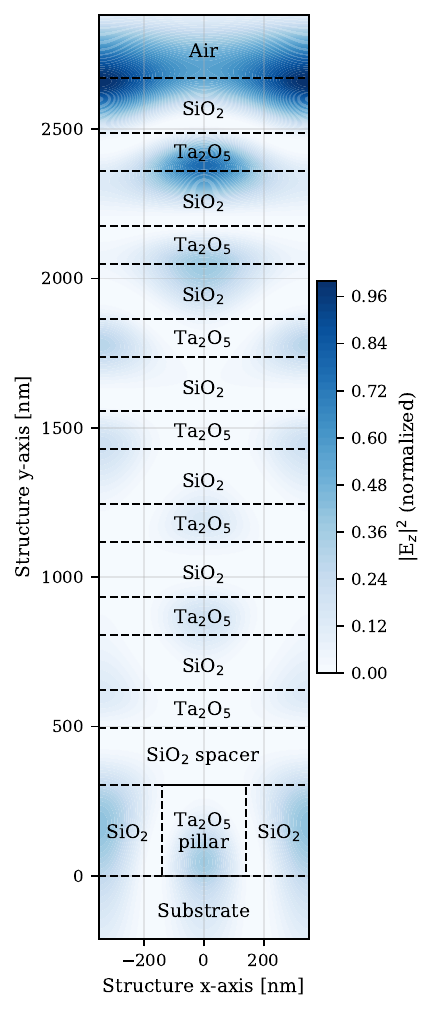}
    \caption{Electric field profile of the TE field for a single unit of the periodic one-dimensional array at resonant wavelength \qty{1064}{\nm}. The field comes in from the top, interacting with the Bragg stack first, followed by the metasurface.}
    \label{fig:mode}
\end{figure}

To further illustrate the mechanism of our proposed hybrid design, we show in \cref{fig:mode} the normalized electric field profile of the TE field for a single unit of the one-dimensional array. The field enters from the top and is largely reflected by the 14-layer Bragg stack. A reduced fraction of the field interacts with the metasurface, consistent with the reduced absorption of the hybrid design shown in \cref{fig:spectra}, and is then resonantly reflected, with the resonant mode visible in the metasurface layer.

\section{Fabrication outline}
\label{sec:fab}

For the embedded metasurface, shown on the left side of \cref{fig:design}, we propose first using electron-beam lithography (EBL) and atomic layer deposition (ALD) to fabricate the Ta$_2$O$_5$ rectangular pillar array, in a process analogous to other metasurface fabrication methods for amorphous oxides~\cite{Devlin:2016}. The surrounding SiO$_2$ can then be deposited via ALD or chemical vapor deposition (CVD). ALD offers better conformality for higher aspect ratio depositions, and better tolerancing of the total layer height, matching the pillar height $h_0$ required by the design, but at a significant time cost relative to CVD. The over-deposited SiO$_2$ is then etched back to planarize the surface at the pillar height $h_0$, yielding the embedded metasurface. Significant process development is required to minimize optical loss from both material absorption and scattering, the latter arising from thin-film deposition defects such as voids and pinholes, as well as from fabrication-induced surface roughness. For the un-patterned layers, the intermediate SiO$_2$ spacer and the quarter-wavelength Ta$_2$O$_5$/SiO$_2$ Bragg layers, we propose ion-beam sputtering (IBS), consistent with current gravitational-wave detector coating practice~\cite{Pinard:2017}. A full experimental demonstration of this fabrication process is left to future work.

\section{Coating Brownian noise}
\label{sec:noise}

In this section, we provide a general method for calculating the coating Brownian noise of hybrid mirrors, i.e., those combining metasurfaces and Bragg stacks in various configurations. We illustrate this method by calculating the coating Brownian noise for our specific hybrid design, highlighting its noise advantage. Such calculations have been done for hybrid designs with low coupling between the metasurface and the Bragg stack by incoherently summing the Brownian noise contributions for each of them~\cite{Dickmann:2018, Kranhold:2026}. For our mirror, a holistic and coherent calculation is needed, due to constructive interference between the two components leading to strong coupling. The method outlined in this section is generally applicable to any level of coupling between the components.

\begin{figure}[]
    \centering
    \includegraphics[width=\columnwidth]{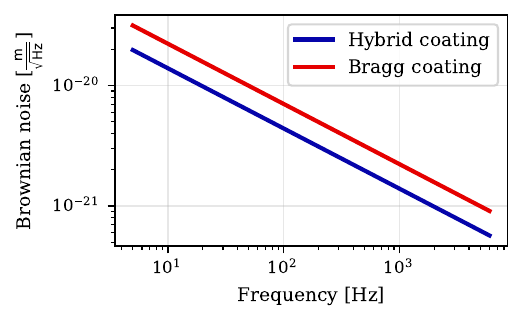}
    \caption{Coating Brownian noise amplitude spectral density as a function of frequency for the proposed hybrid design (blue) and an equivalent-reflectance Bragg stack (red).}
    \label{fig:brownian}
\end{figure}

To calculate the Brownian noise of our designed hybrid mirror, we apply the direct Levin approach~\cite{Levin:1998}, using the electromagnetic mode for a single unit of the periodic array shown in \cref{fig:mode} and obtained using RCWA, to calculate virtual pressure profiles at each interface of our structure from the Maxwell stress tensor. Since the period of our structure $\Lambda_0$ is much smaller than the incident beam radius $r_0$, each such pressure profile contains two physically distinct components: a period-averaged part that varies over $r_0$, and a zero-mean, period-scale remainder that varies over $\Lambda_0$. A periodic unit cell, treated with finite element analysis, captures the elastic response to the period-scale component, as shown in Ref.~\cite{Kroker:2017}. To calculate the response to the beam-scale component requires the infinite half-space treatment of Ref.~\cite{Hong:2013}. We therefore obtain the total coating Brownian noise power spectral density (PSD) by decomposing it into three coherently combined contributions,

\begin{equation}
\label{eq:sdecomp_main}
    S_{yy}(f) = S_{yy}^{(1)}(f) + S_{yy}^{(2)}(f) + S_{yy}^{(\mathrm{cross})}(f),
\end{equation}

The first, $S_{yy}^{(1)}(f)$, arises from the mean-field force varying at the $r_0$ length scale, and is obtained by applying the coating Brownian noise formula of Ref.~\cite{Hong:2013} to our complete layer stack composed of seven Bragg pairs, the intermediate layer, and a homogenized metasurface layer. The second is a periodic term $S_{yy}^{(2)}(f)$, calculated using the exact strain energy density obtained from finite-element analysis of a unit cell with periodic boundary conditions under the periodic pressure component as outlined in Ref.~\cite{Kroker:2017}, followed by a Gaussian-beam array integration. Finally, there is a cross term $S_{yy}^{(\mathrm{cross})}(f)$ arising from the coherent overlap of the two fields. The full calculation is detailed in the Appendix.

The resulting projected amplitude spectral density (ASD) of the coating Brownian noise is plotted in \cref{fig:brownian}, alongside the computed Brownian noise for an equivalent-reflectance Bragg stack~\cite{Hong:2013,gwinc} composed of the same materials, Ta$_2$O$_5$ and SiO$_2$. We take the incident beam radius to be $r_0=$\qty{62}{\mm}, consistent with the values used in Advanced LIGO. The material properties used are tabulated in \cref{tab:material}. This corresponds to a reduction in Brownian noise ASD by a factor of 1.6 relative to the equivalent-reflectance Bragg stack. The mean-field term $S_{yy}^{(1)}(f)$ dominates the total, accounting for 96.23\% of the PSD at 100\,Hz, with the periodic and cross terms contributing 2.24\% and 1.53\%, respectively. The beam-scale mean-field response is the primary contributor to Brownian noise for our design due to our choice of the seven Bragg pairs. Reducing the number of Bragg pairs will increase the contribution from the periodic and cross terms, and thus correctly accounting for them is essential for a rigorous treatment of Brownian noise in hybrid structures.
\section{Angular sensitivity}
\label{sec:sensitivity}

\begin{figure}[tbp]
    \centering
    \includegraphics[width=\columnwidth]{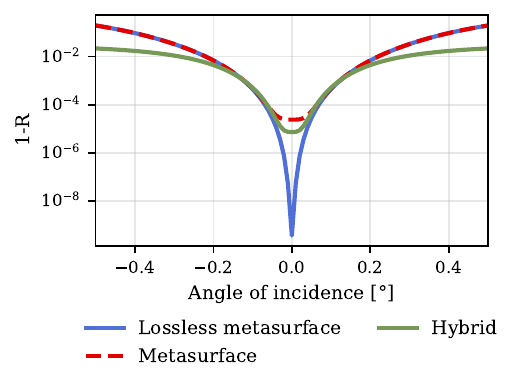}
    \caption{Total loss ($1-R$, where $R$ is reflectance) versus angle of incidence at the resonant wavelength \qty{1064}{\nm}, over a \qty{1}{\degree} range, for the lossless (blue) and lossy (red) bare embedded metasurface compared to the hybrid design (green).}
    \label{fig:angle}
\end{figure}

Finally, we consider the angular sensitivity of our proposed device. Metasurface mirrors, due to their resonant mechanism, are known to be extremely angle-sensitive~\cite{Wang:1990} compared to Bragg mirrors. We examine this by varying the angle of incidence in the plane of the periodic array, i.e., the $xy$ plane in \cref{fig:design}, at the resonant wavelength \qty{1064}{\nm}. We plot the total loss, $1-R$, over an angular range of \qty{1}{\degree} for both the lossless and lossy bare embedded metasurface as specified in \cref{sec:design}, and compare it to that of the hybrid design, as shown in \cref{fig:angle}. Near normal incidence, all designs demonstrate their total loss on resonance. At larger angles of incidence, their behavior diverges: the loss for both bare metasurface designs rises steeply as the resonance condition is no longer satisfied, while the hybrid design's loss remains substantially lower and flattens out, as the response becomes increasingly governed by the Bragg stack's inherently broader angular acceptance. This improved angular tolerance is an important consideration for incorporating such mirrors into practical, large-scale experiments.

\section{Conclusion}
\label{sec:conclusion}

We have proposed a new type of low coating Brownian noise mirror that combines a metasurface and a Bragg stack in a configuration that also minimizes optical loss. We have done so by designing an embedded metasurface that can be placed underneath a Bragg mirror using an intermediate phase-matching layer, so as to preserve the high reflectance and low Brownian noise benefits of a metasurface mirror while minimizing the fraction of incident light that interacts with the more loss-prone metasurface layer. Unlike previous hybrid designs, which place the metasurface at the input surface and thus expose it to the full incident field, our under-Bragg ordering directly reduces this exposure and hence the overall optical loss. We have demonstrated in simulation that this reduces the optical loss by over a factor of 3, from \qty{24}{ppm} for a bare metasurface to \qty{7}{ppm} for our hybrid design, for our specific example considering absorption loss. We expect this reduction to translate qualitatively to scatter loss as well, though quantitative modeling of scatter loss is left to future work. The designed mirror also achieves high reflectance, $R$, ($1-R=$\qty{7}{ppm}) using only \qty{2.7}{\micro\meter} of coating thickness, compared to \qty{5.8}{\micro\meter} for an equivalent-reflectance Bragg stack. We emphasize that the specific parameters used here, including the choice of 7 Bragg pairs, are illustrative of this architecture rather than a claimed optimum, and the same design principles are directly applicable to other operating wavelengths and material systems.

We have also provided a general and rigorous method to calculate the coating Brownian noise of hybrid metasurface-multilayer mirrors, extending prior approaches for periodic metasurface structures and for multilayer coatings to coherently combine their respective contributions. Using this method, we have shown that our hybrid design achieves a reduction in coating Brownian noise by a factor of 1.6 relative to an equivalent-reflectance Bragg stack. This improvement is of interest to current precision measurement experiments, as coating Brownian noise already limits the stability of optical atomic clocks~\cite{Ludlow:2015} and the sensitivity of ground-based gravitational-wave detectors in their mid-frequency band~\cite{Capote:2025}, and a reduction of this magnitude could translate directly into improved sensitivity for both. Finally, we have shown that the hybrid design also has greater angular tolerance than a bare metasurface mirror.

While we have outlined a possible pathway for fabricating the designed mirror, we anticipate many process development steps before achieving this fabrication, and scaling up to large-area mirrors will be essential for incorporating such mirrors into large-scale experiments, in particular gravitational-wave detectors. In this paper, we have considered a single polarization and a one-dimensional metasurface array; extending this design to a two-dimensional array would enable polarization-insensitive operation. Whether this is desirable depends on the specific application, as some experiments may instead benefit from the polarization selectivity of the one-dimensional design presented here.

Coating Brownian noise remains one of the central obstacles to further improving the sensitivity of precision optical measurements. The hybrid architecture presented here offers a practical route towards incorporating metasurface mirrors into real experiments, by addressing the optical loss that has so far limited their use, while the Brownian noise calculation method provides the tools needed to evaluate such hybrid designs rigorously.

\sechead{Acknowledgments}
SP thanks Kevin Kuns for helpful discussions on Brownian noise calculations, and Liyuan Zhang for support with absorption measurements. This work was partially supported by the LIGO Laboratory, which is funded by the U.S. National Science Foundation award PHY-2309200.


\bibliography{mm.bib}

\appendix

\section{Coating Brownian noise}
\label{app:brownian}

As outlined in the main text, we calculate the Brownian noise for our proposed hybrid structure by applying the direct Levin approach~\cite{Levin:1998,Hong:2013}, modified for periodic nanostructures~\cite{Kroker:2017}, to our full combined metasurface-Bragg stack. First, we use the field profile obtained from rigorous coupled-wave analysis (RCWA) simulations to calculate the electromagnetic pressure at the Ta$_2$O$_5$--SiO$_2$ interfaces and the top-most SiO$_2$--air interface for a single unit of the periodic structure. This pressure is given by the difference of the Maxwell stress tensor at the two sides of each interface. Since all materials considered here are non-magnetic ($\mu_\text{r}=1$), the magnetic field contributes no pressure difference across any interface, and the pressure depends only on the electric field. For example, along the top-most SiO$_2$--air interface, which lies in the $zx$ plane, the electromagnetic pressure is given by
\begin{equation}
  P_{\mathrm{top}}(x) = \frac{1}{2}\epsilon_0(\epsilon_\text{S}-1)E_{z,\mathrm{top}}^2(x)
\end{equation}
where $\epsilon_0$ is the vacuum permittivity, $\epsilon_\text{S}$ is the relative permittivity of SiO$_2$, and $E_{z,\mathrm{top}}(x)$ is the electric field in the $\hat{z}$ direction along the top-most interface as a function of direction $\hat{x}$. The structure is periodic in the $\hat{x}$ direction, the light is incoming in the $-\hat{y}$ direction, and polarized in the $\hat{z}$ direction. At each Ta$_2$O$_5$--SiO$_2$ interface $k$, the pressure $P_k(x)$ takes the same form with $(\epsilon_\text{S} - 1)$ replaced by $\pm(\epsilon_\text{T} - \epsilon_\text{S})$, where $\epsilon_\text{T}$ is the relative permittivity of Ta$_2$O$_5$ and the sign depends on the interface order.
(For reasons described later, the interfaces lying outside the $zx$ plane do not need to be considered in our particular geometry.)

\subsection*{Decomposition}

The period $\Lambda_0$ of our structure is sub-wavelength, and hence much smaller than the beam radius $r_0$ of the incident Gaussian beam. This scale separation means that each pressure profile $P_k(x)$ at the $k$\textsuperscript{th} interface contains two physically distinct components: a period-averaged part $\bar{P}_k$, which varies only on the beam scale $r_0$ along the mirror, and a zero-mean, period-scale remainder $\delta P_k(x)$:
\begin{equation}
\label{eq:pdecomp1}
    P_k(x) = \bar P_k + \delta P_k(x),
\end{equation}
where
\begin{equation}
\label{eq:pdecomp2}
    \bar P_k \equiv \frac{1}{\Lambda_0}\int\limits_0^{\Lambda_0} P_k(x)\,dx, \qquad
    \int\limits_0^{\Lambda_0} \delta P_k(x)\,dx = 0.
\end{equation}
A single unit cell of the periodic structure, treated with periodic boundary conditions using finite-element analysis (FEA), correctly captures the elastic response to the period-scale component $\delta P_k(x)$. However, it cannot capture the elastic response to the beam-scale component $\bar{P}_k$, which must be computed over the entire mirror domain (e.g., the infinite half-space treatment for a finite-width beam as given in Ref.~\cite{Hong:2013}). We therefore treat the two components separately for a complete calculation of the Brownian noise of the system. The pillar sidewalls, lying in the $yz$ plane and having no analog in a Bragg stack, are treated entirely as a periodic $\delta P_k(y)$ part.

We consider linear elastic materials, and thus the strain $s_{ij}$ and stress $t_{ij}$ that develop in response to $P_k(x)$ decompose in the same manner:
\begin{align}
    &s_{ij}(\vec{r}) = s^{(1)}_{ij}(\vec{r}) + s^{(2)}_{ij}(\vec{r})\\
    &t_{ij}(\vec{r}) = t^{(1)}_{ij}(\vec{r}) + t^{(2)}_{ij}(\vec{r}),
\end{align}
where $i,j \in \{x,y,z\}$ and $\vec{r} = \begin{pmatrix} x & y &z \end{pmatrix}^T$ denotes the three-dimensional position within the domain. The superscript $(1)$ denotes the field sourced by the mean-pressure components $\{\bar P_k\}$ and $(2)$ the field sourced by the oscillating components $\{\delta P_k(x)\}$.
The stress and strain are related by $t_{ij}(\vec{r}) = C_{ijkl}(\vec{r})\,s_{kl}(\vec{r})$, with $C_{ijkl}(\vec{r})$ the true local elastic stiffness tensor, and repeated spatial indices are summed over.
The stiffness tensor has symmetry $C_{ijkl}(\vec{r}) = C_{klij}(\vec{r})$.

We are interested in the strain energy density $\mathcal{E}(\vec{r}) = \frac{1}{2}s_{ij}(\vec{r})C_{ijkl}(\vec{r})s_{kl}(\vec{r})$, and it can also be decomposed:
\begin{equation}
  \mathcal{E}(\vec{r}) = \mathcal{E}^{(1)}(\vec{r}) + \mathcal{E}^{(2)}(\vec{r}) + \mathcal{E}^{(\mathrm{cross})}(\vec{r}),
  \label{eq:Edecomp}
\end{equation}
with
\begin{subequations}
\begin{align}
  \mathcal{E}^{(1)}(\vec{r}) &= \tfrac{1}{2}s^{(1)}_{ij}(\vec{r})\, C_{ijkl}(\vec{r})\, s^{(1)}_{kl}(\vec{r}), \\
  \mathcal{E}^{(2)}(\vec{r}) &= \tfrac{1}{2}s^{(2)}_{ij}(\vec{r})\, C_{ijkl}(\vec{r})\, s^{(2)}_{kl}(\vec{r}), \\
  \mathcal{E}^{(\mathrm{cross})}(\vec{r}) &= \tfrac{1}{2}s^{(1)}_{ij}(\vec{r})\,C_{ijkl}(\vec{r})\,s^{(2)}_{kl}(\vec{r})\nonumber\\
    &\hphantom{=} {} \qquad + \tfrac{1}{2}s^{(2)}_{ij}(\vec{r})\,C_{ijkl}(\vec{r})\,s^{(1)}_{kl}(\vec{r}).
\end{align}
  \label{eq:Edecomp terms}
\end{subequations}

In the direct method of computing the Brownian noise, the energy density $\mathcal{E}$ is used to compute $W$, the total energy dissipated in response to the application of the total electromagnetic force $F$; the dissipation arises due to the nonzero mechanical loss angle $\phi$ in the silica and tantala.
(We have not included any additional mechanical loss due to the interfaces, as previous studies of ion-beam-sputtered silica--tantala Bragg stacks have shown this loss to be negligible~\cite{Penn:2003nh}.)
The ratio $W / F^2$ determines the power spectral density (PSD) of the Brownian noise as a function of Fourier frequency $f$~%
\footnote{In the direct approach, one needs $W(f)$, the energy dissipated over the period $1/f$ for a single cycle of an applied oscillatory force $F(f)$. We make the assumption that we are interested in Fourier frequencies smaller than any mechanical resonances in the mirror, in which case the elastostatic result at $f = 0$ can be used to infer the entire noise spectrum.}.
We denote this PSD by $S_{yy}(f)$ to indicate it is given as an equivalent displacement fluctuation of the coating surface as sensed by the Gaussian optical beam upon reflection.
Our decomposition of $\mathcal{E}$, and hence $W$, extends to the PSD:
\begin{equation}
\label{eq:Sdecomp}
    S_{yy}(f) = S_{yy}^{(1)}(f) + S_{yy}^{(2)}(f) + S_{yy}^{(\text{cross})}(f),
\end{equation}
where $S_{yy}^{(1)}(f)$ is the mean-field contribution, exactly as calculated in Ref.~\cite{Hong:2013}, and expanded upon in \cref{app:mean}. $S_{yy}^{(2)}(f)$ is the oscillatory term contribution, following the approach outlined in Ref.~\cite{Kroker:2017}, and integrating over a Gaussian incident beam, as expanded upon in \cref{app:periodic}. $S_{yy}^{(\mathrm{cross})}(f)$ is the cross term, calculated using a combination of the oscillatory term and the mean term, expanded upon in \cref{app:cross}.

\begin{figure}[]
    \centering
    \includegraphics[width=\columnwidth]{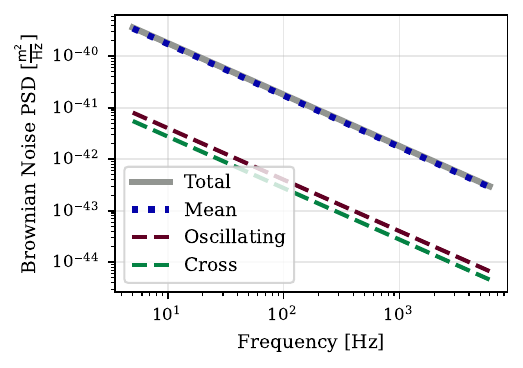}
    \caption{Coating Brownian noise power spectral density contributions from the mean-field term, oscillating periodic term, and cross term of \cref{eq:Sdecomp}, together with their total coherent sum.}
    \label{fig:contributions}
\end{figure}

The three contributions to \cref{eq:Sdecomp} are plotted individually in \cref{fig:contributions} for our illustrative design. We find that the contribution from the mean term $S_{yy}^{(1)}(f)$ accounts for 96.23\% of the total in power spectral density, whereas the oscillating periodic term $S_{yy}^{(2)}(f)$ contributes 2.24\% and the cross term $S_{yy}^{(\mathrm{cross})}(f)$ contributes 1.53\%. We have used 7 Bragg pairs for our example; increasing the number of pairs will decrease the contributions from the periodic and cross terms, and vice versa.

The next subsections compute the total applied electromagnetic force $F_{0,\text{tot}}$, and then the three contributions to the dissipated energy $W$.

\subsection{Applied electromagnetic force}
\label{app:force}

To compute the total applied electromagnetic force, we first compute the total force per unit length in the $\hat{z}$-direction inside the cell and at $z = 0$.
We denote this quantity by $\Tilde{F}_{\mathrm{0,cell}}$.
It is the line integral of the pressure profiles along all the interfaces:
\begin{equation}
\label{eq:funit}
  \Tilde{F}_{0,\text{cell}} = \int\limits_{\text{cell}} P(s)\, ds
    = \int\limits_{\text{cell}} \bar P(s)\, ds,
\end{equation}
where $s$ is the coordinate along each interface: $x$ for the horizontal Ta$_2$O$_5$--SiO$_2$ and SiO$_2$--air interfaces, and $y$ for the vertical sidewalls of the pillar. By symmetry, the contributions from the pillar sidewalls cancel, but are retained here for a complete line integral.

Next, we calculate the total applied force for a Gaussian beam of radius $r_0$ incident on the full periodic array, with power profile
\begin{equation}
  p_{\text{in}}(x,z) \propto \rme^{-2(x^2+z^2)/r_0^2}.
  \label{eq:gauss}
\end{equation}
Since electromagnetic pressure is directly proportional to electric field intensity, it will have the same Gaussian profile. We know that the period $\Lambda_0$, which is sub-wavelength, is much smaller than the beam size. Labeling the central transverse positions of the cells by $x_m = m \Lambda_0$, with $m \in \mathbb{Z}$, we can thus assume that over any single cell, whose transverse extent is $x_m - \Lambda_0 / 2 \le x < x_m + \Lambda_0/2$, the incident beam can be taken to have power $p_\text{in}(x_m,z)$. We also assume that the mirror size is much larger than the incoming beam. Then for the cell located at $x_m$, the total integrated force on a single unit will be given by
\begin{equation}
  F_{0,\text{cell},\text{tot}} \approx \rme^{-2x_m^2/r_0^2} \Tilde{F}_{0,\text{cell}} \left(\,\int\limits_{-\infty}^{\infty} \rme^{-2z^2/r_0^2}dz\right).
  \label{eq:F cell tot}
\end{equation}
Then summing over all the cells, the total integrated force on the entire mirror will be given by
\begin{align}
    F_{0,\text{tot}} &= \Tilde{F}_{0,\mathrm{cell}} \left(\int\limits_{-\infty}^{\infty} \rme^{-2z^2/r_0^2}dz\right) \left( \sum_{m = -\infty}^{\infty} \rme^{-2(m\Lambda_0)^2/r_0^2} \right)\\
    &\approx \Tilde{F}_{0,\text{cell}} \left( \sqrt{\frac{\pi}{2}} r_0 \right) \left( \sqrt{\frac{\pi}{2}} \frac{r_0}{\Lambda_0} \right)
\end{align}
once again using $\Lambda_0 \ll r_0$.

\subsection{Mean-field ($\bar P$) contribution}
\label{app:mean}

The total coating thickness is much smaller than the beam radius $r_0$, and thus the mean pressures $\{\bar P_k\}$ act as if applied at the surface of a coated elastic half-space, under the beam-scale Gaussian envelope with power profile given by \cref{eq:gauss} in the $zx$ plane at $y=0$. This is the geometry treated by Hong~\emph{et al.}~\cite{Hong:2013}; therefore, we can obtain $S^{(1)}(f)$ by applying their full coating Brownian noise formula [Eq.~(94) of Ref.~\cite{Hong:2013}] to an arbitrary, not necessarily alternating, layer stack, required here to accommodate the embedded-metasurface layer, the SiO$_2$ phase-matching layer, and the Bragg doublets within a single calculation. The embedded-metasurface layer is assigned effective mechanical properties (Young's modulus, Poisson's ratio, loss angle) obtained from the Ta$_2$O$_5$/SiO$_2$ fill ratio and the material values in \cref{tab:material}.

\subsection{Periodic ($\delta P$) contribution}
\label{app:periodic}

To find $S_{yy}^{(2)}(f)$, we first apply the zero-mean pressure profiles $\{\delta P_k(x')\}$ to a single cell of the mirror using FEA with periodic boundary conditions, implemented in COMSOL~\cite{comsol}, which yields the resulting strain and stress fields.
Here we have denoted by $x'$ the microscopic transverse position running from $x_m - \Lambda_0/2$ to $x_m + \Lambda_0/2$ for the cell centered at $x_m = m\Lambda_0$.
The transverse strain associated with the $\hat{z}$ direction can only arise from the $\hat{z}$-gradient of the beam envelope, which varies on the beam scale $r_0$. Since this scale is far larger than the periodic structure's own length scale $\Lambda_0$, this contribution is negligible compared to the strain in $\hat{x}$ and $\hat{y}$, which is driven directly by the periodic pressure component. This is distinct from the strain due to the mean-field term, which varies over $r_0$ in both lateral directions.
Therefore, it is sufficient to consider the strain energy density $\mathcal{E}^{(2)}(x',y,0) = \frac{1}{2} t^{(2)}_{ij}(x',y,0)s^{(2)}_{ij}(x',y,0)$ for the cell centered at $x_m=0$, evaluated at $z = 0$ in a two-dimensional geometry.
If the mechanical loss angle at position $(x',y)$ is $\phi(x',y)$, the dissipated energy per unit length in the $\hat{z}$ direction at $z = 0$ over a Fourier period $1/f$ is, for this reference cell,
\begin{equation}
    \tilde W^{(2)}_{\text{cell}} = 2\pi f \int\limits_{\text{cell}} \mathcal{E}^{(2)}(x',y,0)\,\phi(x',y)\,dx'\,dy.
\end{equation}
Since $\mathcal{E}^{(2)}$ is proportional to the square of the local pressure amplitude, its value at the cell centered at $x_m$, and at position $z$, follows the square of the Gaussian profile in \cref{eq:gauss} relative to this reference value, analogous to \cref{eq:F cell tot}: 
\begin{equation}
  W^{(2)}_{\text{cell,tot}}(x_m,z) \approx \rme^{-4x_m^2/r_0^2}\,\tilde W^{(2)}_{\text{cell}} \int\limits_{-\infty}^{+\infty} \rme^{-4z^2/r_0^2} \, dz.
\end{equation}
Summing over all cells, in a manner similar to the computation of the force, the total dissipated energy is then given by
\begin{align}
    W^{(2)}_{\mathrm{tot}} = \Tilde{W}^{(2)}_{\text{cell}} \left(\int\limits_{-\infty}^{\infty} \rme^{-4z^2/r_0^2}dz\right)\left(\sum_{m=-\infty}^{\infty}\rme^{-4(m\Lambda_0)^2/r_0^2}\right)\\ \approx \Tilde{W}^{(2)}_{\text{cell}} \left( \sqrt{\frac{\pi}{4}} r_0 \right) \left( \sqrt{\frac{\pi}{4}} \frac{r_0}{\Lambda_0} \right).
\end{align}

Substituting $W^{(2)}_{\text{tot}}$ and $F_{0,\text{tot}}$ into the fluctuation-dissipation theorem, we obtain the following expression for the power spectral density:
\begin{equation}
    S_{yy}^{(2)}(f) = \frac{2 k_B T}{\pi^2 f^2}\frac{\Tilde{W}^{(2)}_{\mathrm{cell}}}{\Tilde{F}_{0,\text{cell}}^2} \frac{\Lambda_0}{\pi r_0^2}.
\end{equation}

\subsection{Cross term}
\label{app:cross}

$S_{yy}^{(1)}(f)$ was calculated directly from the aggregate formula of Ref.~\cite{Hong:2013} without requiring the underlying local fields. To calculate the cross term $S_{yy}^{(\mathrm{cross})}(f)$, we require the field $s^{(1)}_{ij}(\vec{r})$ explicitly, so that it can be combined pointwise with the periodic field $t^{(2)}_{ij}(\vec{r})$.

From \cref{eq:Edecomp}, using the symmetry $C_{ijkl}(\vec{r})=C_{klij}(\vec{r})$, the cross-term energy density is
\begin{equation}
    \mathcal{E}^{(\mathrm{cross})}(\vec{r}) = s^{(1)}_{ij}(\vec{r})\,C_{ijkl}(\vec{r})\,s^{(2)}_{kl}(\vec{r}) = s^{(1)}_{ij}(\vec{r})\,t^{(2)}_{ij}(\vec{r}).
\end{equation}

Since $\Lambda_0\ll r_0$, $s^{(1)}_{ij}(\vec{r})$ varies only on the beam scale and is constant across any single cell of the periodic array centered at $x=m\Lambda_0$, while $t^{(2)}_{ij}(\vec{r})$, calculated via the local periodic pressure $\delta P_k(x')$, inherits the same Gaussian envelope as the pressure itself. Writing $\bar P_k(x,z) \equiv \bar P_k\,\rme^{-2(x^2+z^2)/r_0^2}$ for the beam-scale envelope of each mean interface pressure [\cref{eq:gauss}], and defining the $\phi$-weighted per-cell stress integral at $x_m=0$, $z=0$,
\begin{equation}
\label{eq:Jij}
    J_{ij}(y) \equiv \int\limits_{\mathrm{cell}} t^{(2)}_{ij}(x',y)\,\phi(x',y)\,dx',
\end{equation}
the value of this integral at the cell centered at $x_m$, $z$, therefore rescales by the same Gaussian factor. The total cross-term dissipated energy is then
\begin{align}
\label{eq:Wcross}
    W^{(\mathrm{cross})}_{\mathrm{tot}} = 2\pi f \int dy \sum_{m=-\infty}^{\infty}\int\limits_{-\infty}^{\infty} dz\; &s^{(1)}_{ij}(m\Lambda_0,y,z)\,J_{ij}(y)\nonumber\\&\times \rme^{-2\left[(m\Lambda_0)^2+z^2\right]/r_0^2}.
\end{align}

The field $s^{(1)}_{ij}(\vec{r})$ is obtained from the coated-half-space solution of Ref.~\cite{Hong:2013}, Appendix~B. Using the coordinate correspondence between Hong \emph{et al.}'s depth axis and our $y$, and their two lateral axes and our $\{x,z\}$, their solution [Eqs.~(B25)--(B27)] gives, for the three lateral strain components,
\begin{multline}
    s^{(1)}_{ab}(x,z) = \frac{(2\nu_\text{s}-1)(1+\nu_\text{s})}{Y_\text{s}} \\
    \times \iint\frac{dk_x\,dk_z}{(2\pi)^2}\,\hat F(k_x,k_z) \frac{k_a k_b}{\kappa^2}\,\rme^{-\rmi(k_x x+k_z z)}
\label{eq:s1lat}
\end{multline}
where $(a,b)\in\{(x,x),(z,z),(x,z)\}$, $\kappa^2=k_x^2+k_z^2$, and $\hat F(k_x,k_z)$ is the two-dimensional Fourier transform of $\sum_k\bar P_k(x,z)$. \cref{eq:s1lat} is evaluated numerically. These strains are independent of $y$. Hong \emph{et al.}'s solution contains no further nonzero lateral strain components at this order: $s^{(1)}_{xy}\equiv0$ and $s^{(1)}_{yz}\equiv0$ identically, for any lateral pressure profile, since a purely normal applied load produces no lateral-depth shear response in the coated-half-space geometry.

By mechanical equilibrium, the vertical stress at the beam center at height $y$ is
\begin{equation}
  t^{(1)}_{yy}(y) = -\sum_{y_k > y} \bar{P}_k
  \label{eq:t1yy}
\end{equation}
summed over all interfaces $k$ lying above height $y$. Since every interface pressure shares the same lateral envelope, this generalizes to arbitrary lateral position as $t^{(1)}_{yy}(\vec{r}) = \rme^{-2(x^2+z^2)/r_0^2}\,t^{(1)}_{yy}(y)$. At each height $y$, the vertical strain then follows directly from the isotropic form of Hooke's law, using $s^{(1)}_{xx}$ and $s^{(1)}_{zz}$ [from \cref{eq:s1lat}] and $t^{(1)}_{yy}(\vec{r})$ (as given above):
\begin{multline}
    s^{(1)}_{yy}(\vec{r}) = \frac{[1+\nu(y)][1-2\nu(y)]}{Y(y)[1-\nu(y)]}\,t^{(1)}_{yy}(\vec{r})\\
    {}- \frac{\nu(y)}{1-\nu(y)}\left[s^{(1)}_{xx}(x,z)+s^{(1)}_{zz}(x,z)\right]
\label{eq:s1yy}
\end{multline}
which uses the local material properties $Y(y)$ and $\nu(y)$ of whichever layer occupies height $y$, with the homogenized effective properties used within the embedded-metasurface layer as in \cref{app:mean}.

Every term of $s^{(1)}_{ij}(\vec{r})$ is now separable into a product of a depth profile and a lateral profile: the lateral strains of \cref{eq:s1lat} are independent of $y$, while both terms of \cref{eq:s1yy} factor into a $y$-dependent material profile times a lateral field. The integrals over depth (microscopic) and lateral position (macroscopic) in \cref{eq:Wcross} therefore fully separate. We define the lateral beam-scale integrals
\begin{equation}
\label{eq:Gab}
    G_{ab} \equiv \sum_{m=-\infty}^{\infty}\int\limits_{-\infty}^{\infty} dz\; s^{(1)}_{ab}(m\Lambda_0,z)\,e^{-2\left[(m\Lambda_0)^2+z^2\right]/r_0^2}
\end{equation}
for $ab\in\{xx,zz,xz\}$, and the depth integrals
\begin{align}
\label{eq:Kab}
    &K_{ab} \equiv \int dy\,J_{ab}(y), \qquad ab\in\{xx,zz\},\nonumber\\
    &K^{t}_{yy} \equiv \int dy\,\frac{(1+\nu(y))(1-2\nu(y))}{Y(y)(1-\nu(y))}\,t^{(1)}_{yy}(y)\,J_{yy}(y),\nonumber\\
    &K^{s}_{yy} \equiv \int dy\,\frac{\nu(y)}{1-\nu(y)}\,J_{yy}(y),
\end{align}
where $J_{yy}$ requires two depth integrals because the two terms of \cref{eq:s1yy} carry different lateral profiles. The $s^{(1)}_{xz}J_{xz}$ contribution vanishes identically: the kernel $k_xk_z/\kappa^2$ in \cref{eq:s1lat} is odd under $k_z\to-k_z$ while $\hat F$ is even for the circularly symmetric beam of \cref{eq:gauss}, so $s^{(1)}_{xz}(x,z)$ is odd in $z$ and its lateral integral $G_{xz}$ vanishes for any periodic structure. For the remaining lateral integral of the bare Gaussian envelope, appearing with the $t^{(1)}_{yy}$ term, the identical summation as in \cref{app:periodic} gives $\pi r_0^2/2\Lambda_0$ in the limit $\Lambda_0\ll r_0$. Substituting into \cref{eq:Wcross} and separating the $y-$ and lateral integrals, the total cross-term dissipated energy becomes
\begin{multline}
  W^{(\mathrm{cross})}_{\mathrm{tot}} = 2\pi f \left[G_{xx}K_{xx} + G_{zz}K_{zz} + \frac{\pi r_0^2}{2\Lambda_0}K^{t}_{yy} \right. \\
  \left. \vphantom{\frac{r_0^2}{2}} - \left(G_{xx}+G_{zz}\right)K^{s}_{yy}\right]
  \label{eq:Wcross_sep}
\end{multline}

Substituting $W^{(\mathrm{cross})}_{\mathrm{tot}}$ from \cref{eq:Wcross_sep} and $F_{0,\mathrm{tot}}$ from \cref{app:force} into the fluctuation-dissipation theorem gives
\begin{equation}
    S_{yy}^{(\mathrm{cross})}(f) = \frac{2k_BT}{\pi^2f^2}\frac{W^{(\mathrm{cross})}_{\mathrm{tot}}}{F_{0,\mathrm{tot}}^2}.
\end{equation}

\section{Extinction coefficient estimation}

The extinction coefficient values used for atomic layer deposition (ALD) based Ta$_2$O$_5$ and SiO$_2$ in \cref{tab:material} in the main text are extrapolated from our own measurements of ALD-deposited TiO$_2$:HfO$_2$ nanolaminate films, rather than direct measurements for Ta$_2$O$_5$ or SiO$_2$, which we were unable to find in the ALD literature. Using photothermal common-path interferometry (PCI) at \qty{1064}{\nm}, we measured an extinction coefficient of $k\approx10^{-6}$ for our TiO$_2$:HfO$_2$ film following a post-deposition anneal. This was the lowest value obtained across a range of deposition and annealing conditions, reflecting the expected outcome of further process development for our proposed fabrication approach. We take this as an order-of-magnitude estimate for ALD-deposited Ta$_2$O$_5$, on the basis that both are amorphous oxides deposited via ALD under comparable process conditions. For SiO$_2$, we assume an extinction coefficient an order of magnitude lower than Ta$_2$O$_5$, consistent with the analogous trend observed between IBS-deposited SiO$_2$ and Ta$_2$O$_5$ in current gravitational-wave detector coatings~\cite{Granata:2020}.

\end{document}